\documentclass[a4paper,10pt]{article}
\pdfoutput=1
\usepackage[numbers]{natbib}
\usepackage[margin=1in]{geometry}

\usepackage{amssymb}
\usepackage[cmex10]{amsmath}
\usepackage{amsfonts}
\usepackage{array,arydshln}
\usepackage{algorithmic}
\usepackage{algorithm}
\usepackage{etex}
\usepackage{multirow}

\usepackage{tabularx, booktabs}
\newcolumntype{Y}{>{\centering\arraybackslash}X}

\newcommand{\Transp}{\mathsf{T}}

\DeclareMathOperator*\argmin{arg \, min \,}
\DeclareMathOperator*\argmax{arg \, max \,}
\DeclareMathOperator{\Tr}{Tr}
\DeclareMathOperator*\vect{vec}

\newcommand{\SO}[1]{SO(#1)}
\usepackage{booktabs}
\usepackage{enumitem}

\usepackage{amsthm}

\theoremstyle{assumption}
\newtheorem{assumption}{Assumption}[]

\usepackage{tikz,pgfplots}
\usepgfplotslibrary{external}
\pgfplotsset{compat=newest}
\usepackage{pgfplotstable}
\usepackage{acronym}

\usepackage[caption=false,font=footnotesize]{subfig}

\graphicspath{{./figures/}}

 \newlength\figureheight 
    \newlength\figurewidth 
\newcommand{\ie}{i.e.\ }
\newcommand{\eg}{e.g.\ }

\newcommand{\dr}{Dr.\ }
\newcommand{\Sectionref}[1]{Section~\ref{#1}}
\newcommand{\Figureref}[1]{Figure~\ref{#1}}
\newcommand{\Tableref}[1]{Table~\ref{#1}}
\newcommand{\Stepref}[1]{Step~\ref{#1}}
\newcommand{\Algorithmref}[1]{Algorithm~\ref{#1}}
\newcommand{\Assumptionref}[1]{Assumption~\ref{#1}}

\acrodef{imu}[IMU]{inertial measurement unit}
\acrodef{ml}[ML]{maximum likelihood}
\acrodef{ekf}[EKF]{extended Kalman filter}
\acrodef{mekf}[MEKF]{multiplicative EKF}
\acrodef{rmse}[RMSE]{root mean square error}
\acrodef{rms}[RMS]{root mean square}

\providecommand{\keywords}[1]{\textbf{\textit{Keywords: }} #1}

\begin{document}

\newcommand{\coverTitle}{Magnetometer calibration using inertial sensors}
\newcommand{\coverAuthors}{Manon~Kok and Thomas~B.~Sch\"on}
\newcommand{\coverYear}{2016}
\newcommand{\coverStatus}{Accepted for publication.}

\begin{titlepage}
\begin{center}
{\large \em Technical report}

\vspace*{2.5cm}
%
{\Huge \bfseries \coverTitle  \\[0.4cm]}

%
{\Large \coverAuthors \\[2cm]}

\renewcommand\labelitemi{\color{red}\large$\bullet$}
\begin{itemize}
\item {\Large \textbf{Please cite this version:}} \\[0.4cm]
\large
\coverAuthors. \coverTitle. \textit{IEEE Sensors Journal},
Volume 16, Issue 14, Pages 5679--5689, 2016. 
\end{itemize}
\vfill

\begin{abstract}
In this work we present a practical algorithm for calibrating a magnetometer for the presence of magnetic disturbances and for magnetometer sensor errors. To allow for combining the magnetometer measurements with inertial measurements for orientation estimation, the algorithm also corrects for misalignment between the magnetometer and the inertial sensor axes. The calibration algorithm is formulated as the solution to a maximum likelihood problem and the computations are performed offline. The algorithm is shown to give good results using data from two different commercially available sensor units. Using the calibrated magnetometer measurements in combination with the inertial sensors to determine the sensor's orientation is shown to lead to significantly improved heading estimates. 
\end{abstract}

\keywords{Magnetometers, calibration, inertial sensors, maximum likelihood, grey-box system identification, sensor fusion.}

\vfill
\end{center}
\end{titlepage}

\title{\textbf{Magnetometer calibration using inertial sensors}}

\author{Manon~Kok$^\star$ and Thomas~B.~Sch\"on$^\dagger$ \\
\small{$^\star$Department of Electrical Engineering, Link\"oping University, SE-581~83~Link\"oping, Sweden} \\
\small{E-mail: manon.kok@liu.se} \\
\small{$^\dagger$Department of Information Technology, Uppsala University, Sweden} \\
\small{E-mail: thomas.schon@it.uu.se} 
}

\maketitle

    \begin{abstract}
                        In this work we present a practical algorithm for calibrating a magnetometer for the presence of magnetic disturbances and for magnetometer sensor errors. To allow for combining the magnetometer measurements with inertial measurements for orientation estimation, the algorithm also corrects for misalignment between the magnetometer and the inertial sensor axes. The calibration algorithm is formulated as the solution to a maximum likelihood problem and the computations are performed offline. The algorithm is shown to give good results using data from two different commercially available sensor units. Using the calibrated magnetometer measurements in combination with the inertial sensors to determine the sensor's orientation is shown to lead to significantly improved heading estimates. 
    \end{abstract}
    
    \keywords{Magnetometers, calibration, inertial sensors, maximum likelihood, grey-box system identification, sensor fusion.}

\section{Introduction}
\label{sec:introduction}
Nowadays, magnetometers and inertial sensors (gyroscopes and accelerometers) are widely available, for instance in dedicated sensor units and in smartphones. Magnetometers measure the local magnetic field. When no magnetic disturbances are present, the magnetometer measures a constant local magnetic field vector. This vector points to the local magnetic north and can hence be used for heading estimation. Gyroscopes measure the angular velocity of the sensor. Integration of the gyroscope measurements gives information about the change in orientation. However, it does not provide absolute orientation estimates. Furthermore, the orientation estimates suffer from integration drift. Accelerometers measure the sensor's acceleration in combination with the earth's gravity. In the case of small or zero acceleration, the measurements are dominated by the gravity component. Hence, they can be used to estimate the inclination of the sensor. 

Inertial sensors and magnetometers have successfully been used to obtain accurate 3D orientation estimates for a wide range of applications. For this, however, it is imperative that the sensors are properly calibrated and that the sensor axes are aligned. Calibration is specifically of concern for the magnetometer, which needs recalibration whenever it is placed in a (magnetically) different environment. When the magnetic disturbance is a result of the mounting of the magnetometer onto a magnetic object, the magnetometer can be calibrated to compensate for the presence of this disturbance. This is the focus of this work.

\begin{figure}[t]
	\centering
	\setlength\figureheight{0.55\columnwidth}
	\setlength\figurewidth{0.55\columnwidth}
	\includegraphics[width = 0.4\columnwidth]{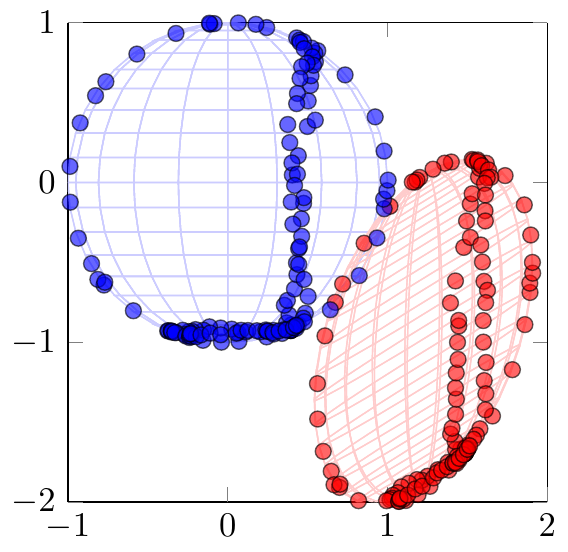}
	\caption{Example calibration results with an ellipsoid of magnetometer data before calibration (red) and a unit sphere of data after calibration (blue).}
	\label{fig:ellipseXsens}
\end{figure}

Our main contribution is a practical magnetometer calibration algorithm that is designed to improve orientation estimates when combining calibrated magnetometer data with inertial data. The word \textit{practical} refers to the fact that the calibration does not require specialized additional equipment and can therefore be performed by any user. More specifically, this means that the orientation of the sensor is not assumed to be known. Instead, the calibration problem is formulated as an orientation estimation problem in the presence of unknown parameters and is posed as a \ac{ml} problem. The algorithm calibrates the magnetometer for the presence of magnetic disturbances, for magnetometer sensor errors and for misalignment between the magnetometer and the inertial sensor axes. Using the calibrated magnetometer measurements to estimate the sensor's orientation is experimentally shown to lead to significantly improved heading estimates. We aggregate and extend the work from~\cite{kokS:2014} and~\cite{kokHSGL:2012} with improvements on the implementation of the algorithm. Furthermore, we include a more complete description and analysis, more experimental results and a simulation study illustrating the heading accuracy that can be obtained with a properly calibrated sensor. 

To perform the calibration, the sensor needs to be rotated in all possible orientations. A perfectly calibrated magnetometer would in that case measure rotated versions of the local magnetic field vector. Hence, the magnetometer data would lie on a sphere. In practice, however, the magnetometer will often measure an ellipsoid of data instead. The calibration maps the ellipsoid of data to a sphere as illustrated in \Figureref{fig:ellipseXsens}. The alignment of the inertial and magnetometer sensor axes determines the orientation of the sphere. Since we are interested in improving the heading estimates, the actual magnitude of the local magnetic field is of no concern. Hence, we assume without loss of generality that the norm is equal to $1$, \ie the sphere in \Figureref{fig:ellipseXsens} is a unit sphere.

\section{Related work}
\label{sec:relatedWork}
Traditional magnetometer calibration approaches assume that a reference sensor is available which is able to provide accurate heading information. A well-known example of this is compass swinging~\cite{bowditch:2002}. To allow for any user to perform the calibration, however, a large number of approaches have been developed that remove the need for a source of orientation information. One class of these magnetometer calibration algorithms focuses on minimizing the difference between the magnitude of the measured magnetic field and that of the local magnetic field, see \eg \cite{alonsoS:2002}. This approach is also referred to as scalar checking~\cite{lerner:1978}. Another class formulates the calibration problem as an ellipsoid fitting problem, \ie as the problem of mapping an ellipsoid of data to a sphere, see \eg \cite{vasconcelosESOC:2011,renaudinAL:2010,gebreEgziabherEPP:2006}. The benefit of using this formulation, is that there is a vast literature on solving ellipsoid fitting problems, see \eg \cite{ganderGS:1994,markovskyKH:2004}. Outside of these two classes, a large number of other calibration approaches is also available, for instance~\cite{wuS:2015}, where different formulations of the calibration problem in terms of an \ac{ml} problem are considered.

The benefit of the approaches discussed above is that they can be used with data from a magnetometer only. Our interest, however, lies in calibrating a magnetometer for improved heading estimation in combination with inertial sensors. Alignment of the sensor axes of the inertial sensors and the magnetometer is in this case crucial. This alignment can be seen as determining the orientation of the blue sphere of calibrated magnetometer data in \Figureref{fig:ellipseXsens}. Algorithms that only use magnetometer data can map the red ellipsoid of data to a sphere, but without additional information, the rotation of this sphere remains unknown.

A number of recent approaches include a second step in the calibration algorithm to determine the misalignment~\cite{vasconcelosESOC:2011,liL:2012, salehiMB:2012, bonnetBGLB:2009} between different sensor axes. A common choice to align the magnetometer and inertial sensor axes, is to use accelerometer measurements from periods of fairly small accelerations~\cite{liL:2012,salehiMB:2012}. The downside of this approach is that a threshold for using accelerometer measurements needs to be determined. Furthermore, data from the gyroscope is hereby omitted. In \cite{troniW:2013} on the other hand, the problem is reformulated in terms of the change in orientation, allowing for direct use of the gyroscope data.

In our algorithm we instead formulate the magnetometer calibration problem as a problem of estimating the sensor's orientation in the presence of unknown (calibration) parameters. This formulation naturally follows from the fact that the problem of orientation estimation and that of magnetometer calibration are inherently connected: If the magnetometer is properly calibrated, good orientation estimates can be obtained. Reversely, if the orientation of the sensor is known accurately, the rotation of the sphere in \Figureref{fig:ellipseXsens} can accurately be determined, resulting in a good magnetometer calibration. In this formulation, data from the accelerometer \emph{and} the gyroscope is used to aid the magnetometer calibration.

Our formulation of the calibration problem requires solving a non-convex optimization problem to obtain \ac{ml} estimates of the calibration parameters. To obtain good initial values of the parameters, an ellipsoid fitting problem and a misalignment estimation problem are solved. Solving the calibration problem as a two-step procedure is similar to the approaches in~\cite{liL:2012,salehiMB:2012}. We analyze the quality of the initial estimates and of the \ac{ml} estimates in terms of their heading accuracy, both for experimental and simulated data. Based on this analysis, we show that significant heading accuracy improvements can be obtained by using the \ac{ml} estimates of the parameters.

\section{Problem formulation}
\label{sec:problemFormulation}
Our magnetometer calibration algorithm is formulated as a problem of determining the sensor's orientation in the presence of unknown model parameters $\theta$. It can hence be considered to be a grey-box system identification problem. A nonlinear state space model on the following form is used
\begin{subequations}
    \label{eq:greyBoxProblem}
  \begin{align}
    x_{t+1} &= f_t(x_t, \omega_t, e_{\omega,t}, \theta) , \label{eq:greyBoxProblemDyn} \\
    y_t &= \begin{pmatrix} y_{\text{a},t} \\ y_{\text{m},t} \end{pmatrix} = \begin{pmatrix} h_{\text{a},t}(x_t) \\ h_{\text{m},t} (x_t, \theta) \end{pmatrix} + e_t(\theta), \label{eq:greyBoxProblemMeas}
  \end{align}
\end{subequations}
where the state $x_t$ represents the sensor's orientation at time~$t$. We use the change in orientation, \ie the angular velocity $\omega_t$, as an input to the dynamic model $f_t(\cdot)$. The angular velocity is measured by the gyroscope. However, the measurements $y_{\omega,t}$ are corrupted by a constant bias $\delta_\omega$ and Gaussian i.i.d. measurement noise with zero mean and covariance $\Sigma_\omega$, \ie $e_{\omega,t} \sim \mathcal{N}(0_{3 \times 1},\Sigma_\omega)$.  

The measurement models $h_{\text{a},t}( \cdot )$ and $h_{\text{m},t}( \cdot )$ in~\eqref{eq:greyBoxProblemMeas} describe the accelerometer measurements $y_{\text{a},t}$ and the magnetometer measurements $y_{\text{m},t}$, respectively. The accelerometer measurement model assumes that the acceleration of the sensor is small compared to the earth gravity. Since the magnetometer is not assumed to be properly calibrated, the magnetometer measurement model $h_{\text{m},t}( \cdot )$ depends on the parameter vector $\theta$. The exact details of the magnetometer measurement model will be introduced in \Sectionref{sec:modeling}. The accelerometer and magnetometer measurements are corrupted by Gaussian i.i.d. measurement noise 
\begin{align}
e_t = \begin{pmatrix} e_{\text{a},t} \\ e_{\text{m},t} \end{pmatrix}\sim \mathcal{N} \left(0_{6 \times 1},\begin{pmatrix} \Sigma_\text{a} & 0_{3 \times 3} \\ 0_{3 \times 3} & \Sigma_\text{m} \end{pmatrix} \right).
\end{align} 

The calibration problem is formulated as an \ac{ml} problem. Hence, the parameters $\theta$ in~\eqref{eq:greyBoxProblem} are found by maximizing the likelihood function $p_{\theta}(y_{1:N})$,
\begin{align}
  \label{eq:ML}
    \widehat{\theta}^{\text{ML}} = \argmax_{\theta\in\Theta}{p_{\theta}(y_{1:N})},
\end{align}
where $y_{1:N} = \{y_1, \dots, y_N\}$ and $\Theta \subseteq\mathbb{R}^{n_{\theta}}$. Using conditional probabilities and the fact that the logarithm is a monotonic function we have the following equivalent formulation of~\eqref{eq:ML},
\begin{align}
  \label{eq:logML}
  \widehat{\theta}^{\text{ML}} = \argmin_{\theta\in\Theta}{-\sum_{t=1}^{N}\log p_{\theta}(y_{t}\mid
    y_{1:t-1})},
\end{align}
where we use the convention that $y_{1:0} \triangleq \emptyset$. The \ac{ml}
estimator~\eqref{eq:logML} enjoys well-understood theoretical properties including strong
consistency, asymptotic normality, and asymptotic efficiency~\cite{ljung:1999}. 

The state space model~\eqref{eq:greyBoxProblem} is nonlinear, implying that there is no closed form solution available for the one
step ahead predictor $p_{\theta}(y_{t}\mid y_{1:t-1})$ in~\eqref{eq:logML}. This can systematically be handled using sequential Monte
Carlo methods (\eg particle filters and particle smoothers), see \eg  \cite{schonWN:2011,
  lindstenS:2013}. However, for the magnetometer calibration problem it is sufficient to
make use of a more pragmatic approach; we simply approximate the one step ahead predictor using an
\ac{ekf}. The result is
\begin{align}
  \label{eq:EKF1pred}
  p_{\theta}(y_t\mid y_{1:t-1}) \approx \mathcal{N}\left(y_t \, ; \, \widehat{y}_{t\mid t-1}(\theta),
    S_t (\theta) \right),
\end{align}
where the mean value $\widehat{y}_{t\mid t-1}(\theta)$ and the covariance~$S_t(\theta)$ are obtained from the \ac{ekf}~\cite{gustafsson:2012}. Inserting~\eqref{eq:EKF1pred}
into~\eqref{eq:logML} and neglecting all constants not depending on $\theta$ results in the following optimization problem,
\begin{align}
  \label{eq:MLapprox}
  \min_{\theta\in\Theta} \frac{1}{2}\sum_{t=1}^{N}\|y_t - \widehat{y}_{t\mid t-1}(\theta)\|_{S_t^{-1}(\theta)}^2 + \log\det S_t (\theta),
\end{align}
which we can solve for the unknown parameters $\theta$. The problem \eqref{eq:MLapprox} is non-convex, implying that a good initial value for $\theta$ is required.

\section{Magnetometer measurement model}
\label{sec:modeling}
In the case of perfect calibration, a magnetometer measures the local magnetic field and its measurements will therefore lie on a sphere with a radius equal to the local magnetic field. Since we are interested in using the magnetometer measurements to improve the orientation estimates from the state space model~\eqref{eq:greyBoxProblem}, the actual magnitude of the local magnetic field is of no concern. Hence, we assume without loss of generality that its norm is equal to one. We denote the normalized local magnetic field by $m^\text{n}$. Ideally, the magnetometer measurements then lie on a sphere with radius equal to one as
\begin{align}
h_{\text{m},t} = m_{t}^\text{b}  = R_t^{\text{bn}} m^\text{n},
\label{eq:rotationEarthMagField}
\end{align}
where $h_{\text{m},t}$ is defined in~\eqref{eq:greyBoxProblemMeas}. The explicit dependence on $x_t$ and $\theta$ has been omitted for notational simplicity. The matrix $R_t^{\text{bn}}$ is the rotation matrix representation of the orientation at time $t$. The superscript $bn$ denotes that the rotation is from the \textit{navigation frame} $n$ to the \textit{body frame} $b$. The body frame $b$ is aligned with the sensor axes. The navigation frame $n$ is aligned with the earth's gravity and the local magnetic field. In case the coordinate frame in which a vector is defined can be ambiguous, we explicitly indicate in which coordinate frame the vector is expressed by adding a superscript $b$ or $n$. Hence, $m^\text{n}$ denotes the normalized local magnetic field in the navigation frame $n$ while $m_t^\text{b}$ denotes the normalized local magnetic field in the body frame $b$. The latter is time-dependent and therefore also has a subscript~$t$. Note that the rotation from navigation frame to body frame is denoted $R_t^{\text{nb}}$ and $R^{\text{bn}}_t = ( R^{\text{nb}}_t )^\Transp$.

In outdoor environments, the local magnetic field is equal to the local \textit{earth} magnetic field. Its horizontal component points towards the earth's magnetic north pole. The ratio between the horizontal and vertical component depends on the location on the earth and can be expressed in terms of the \textit{dip angle}~$\delta$. In indoor environments, the magnetic field can locally be assumed to be constant and points towards a \textit{local} magnetic north. This is not necessarily the earth's magnetic north pole. Choosing the navigation frame $n$ such that the $x$-axis is pointing towards the local magnetic north, ${m}^\text{n}$ can be parametrized in terms of its vertical component $m^\text{n}_z$ 
\begin{subequations}
\label{eq:generalEqDipAngle}
\begin{align}
m^\text{n} &= \begin{pmatrix} \sqrt{1 - \left( m^\text{n}_z \right)^2} & 0 & m^\text{n}_z \end{pmatrix}^\Transp, \label{eq:magfieldVertical} \\
\intertext{or in terms of the dip angle $\delta$}
m^\text{n} &= \begin{pmatrix} \cos \delta & 0 & - \sin \delta \end{pmatrix}^\Transp. \label{eq:magfieldDip}
\end{align}
\end{subequations}%
Note that the two parametrizations do not encode exactly the same knowledge about the magnetic field; the first component of $m^\text{n}$ in~\eqref{eq:magfieldVertical} is positive by construction while this is not true for~\eqref{eq:magfieldDip}. However, both parametrizations will be used in the remainder. It will be argued that no information is lost by using~\eqref{eq:magfieldDip} if the parameter estimates are properly initialized.

The main need for magnetometer calibration arises from the fact that a magnetometer needs recalibration each time it is placed in a magnetically different environment. Specifically, a magnetometer measures a superposition of the local magnetic field and of the magnetic field due to the presence of magnetic material in the vicinity of the sensor. In case this magnetic material is rigidly attached to the magnetometer, it is possible to calibrate the magnetometer measurements for this. The magnetic material can give rise to both hard and soft iron contributions to the magnetic field. Hard iron effects are due to permanent magnetization of the magnetic material and lead to a constant $3 \times 1$ offset vector $o_{\text{hi}}$. Soft iron effects are due to magnetization of the material as a result of an external magnetic field and therefore depend on the orientation of the material with respect to the local magnetic field. We model this in terms of a $3\times3$ matrix $C_{\text{si}}$. Hence, the magnetometer measurements do not lie on a sphere as in~\eqref{eq:rotationEarthMagField}, but instead, they lie on a translated ellipsoid as
\begin{align}
h_{\text{m},t} = C_\text{si} R_t^{\text{bn}} m^\text{n} + o_\text{hi}.
\label{eq:magModel_sh}
\end{align}

As discussed in \Sectionref{sec:relatedWork}, when calibrating the magnetometer to obtain better orientation estimates, it is important that the magnetometer and the inertial sensor axes are aligned. Let us now be more specific about the definition of the body frame $b$ and define it to be located in the center of the accelerometer triad and aligned with the accelerometer sensor axes. Furthermore, let us assume that the accelerometer and gyroscope axes are aligned. Defining the rotation between the body frame $b$ and the magnetometer sensor frame $b_\text{m}$ as $R^{\text{b}_\text{m} \text{b}} $, the model~\eqref{eq:magModel_sh} can be extended to
\begin{align}
h_{\text{m},t} = C_{\text{si}} R^{\text{b}_\text{m} \text{b}} R_t^{\text{bn}} m^\text{n} + o_{\text{hi}} .
\label{eq:magModel_sh_mis}
\end{align}

Finally, the magnetometer calibration can also correct for the presence of sensor errors in the magnetometer. These errors are sensor-specific and can differ for each individual magnetometer. They can be subdivided into three components, see \eg \cite{gebreEgziabherEPP:2006,renaudinAL:2010,vasconcelosESOC:2011}:
\begin{enumerate}
\item Non-orthogonality of the magnetometer axes, represented by a matrix $C_{\text{no}}$.
\item Presence of a zero bias or null shift, implying that the magnetometer will measure a non-zero magnetic field even if the magnetic field is zero, defined by $o_{\text{zb}}$. 
\item Difference in sensitivity of the three magnetometer axes, represented by a diagonal matrix $C_{\text{sc}}$. 
\end{enumerate}
We can therefore extend the model~\eqref{eq:magModel_sh_mis} to also include the magnetometer sensor errors as
\begin{align}
h_{\text{m},t} = C_{\text{sc}} C_{\text{no}} \left( C_{\text{si}} R^{\text{b}_\text{m} b} R_t^{\text{bn}} m^\text{n} + o_{\text{hi}} \right) + o_{\text{zb}}.
\label{eq:magModel_sh_mis_sen}
\end{align}

To obtain a correct calibration, it is fortunately not necessary to identify all individual contributions of the different components in~\eqref{eq:magModel_sh_mis_sen}. Instead, they can be combined into a $3 \times 3$ distortion matrix $D$ and a $3 \times 1$ offset vector ${o}$ where
\begin{subequations}
\begin{align}
D &=  C_{\text{sc}} C_{\text{no}} C_{\text{si}} R^{\text{b}_\text{m} b}, \\
o &= C_{\text{sc}} C_{\text{no}} o_{\text{hi}}+ o_{\text{zb}}.
\end{align}
\label{eq:DoComponents}%
\end{subequations}
The resulting magnetometer measurement model in~\eqref{eq:greyBoxProblemMeas} can be written as
\begin{align}
y_{\text{m},t} = D R_t^{\text{bn}} m^\text{n} + o + e_{\text{m},t}.
\label{eq:modelMagEq}
\end{align}
In deriving the model we have made two important assumptions: 

\begin{assumption}
\label{ass:Do}
The calibration matrix $D$ and offset vector $o$ in~\eqref{eq:DoComponents} are assumed to be time-independent. This implies that we assume that the magnetic distortions are constant and rigidly attached to the sensor. Also, the inertial and the magnetometer sensor axes are assumed to be rigidly attached to each other, \ie their misalignment is represented by a constant rotation matrix. Additionally, in our algorithm we will assume that their misalignment can be described by a rotation matrix, \ie that their axes are not mirrored with respect to each other.
\end{assumption}

\begin{assumption}
\label{ass:mn}
The local magnetic field $m^\text{n}$ is assumed to be constant. In outdoor environments, this is typically a physically reasonable assumption. In indoor environments, however, the local magnetic field can differ in different locations in the building and care should be taken to fulfill the assumption.
\end{assumption}

\begin{algorithm}[t]
\caption{Magnetometer and inertial calibration}
\label{alg:ml}
\begin{enumerate}
\item \label{algstep:initial} Determine an initial parameter estimate $\widehat{D}_0$, $\widehat{o}_0$, $\widehat{m}^\text{n}_0$, $\widehat{\delta}_{\omega,0}$, $\widehat{\Sigma}_{\omega,0}$, $\widehat{\Sigma}_{\text{a},0}$, $\widehat{\Sigma}_{\text{m},0}$ using three steps
\begin{enumerate}
\item Initialize $\widehat{\delta}_{\omega,0}$, $\widehat{\Sigma}_{\omega,0}$, $\widehat{\Sigma}_{\text{a},0}$, $\widehat{\Sigma}_{\text{m},0}$.
\item Obtain an initial $\widetilde{D}_0$ and $\widehat{o}_0$ based on ellipsoid fitting (see \Sectionref{sec:initEllipse}).
\item Obtain initial $\widehat{D}_0$, $\widehat{o}_0$ and $\widehat{m}^\text{n}_0$ by initial determination of the sensor axis misalignment (see \Sectionref{sec:initMisalignment}).
\end{enumerate}
\item \label{algstep:ml} Set $i = 0$ and repeat, 
\begin{enumerate}
\item \label{algstep:ekf} Run the \ac{ekf} using the current estimates $\widehat{D}_i, \widehat{o}_i, \widehat{m}^\text{n}_i$, $\widehat{\delta}_{\omega,i}$, $\widehat{\Sigma}_{\omega,i}$, $\widehat{\Sigma}_{\text{a},i}$, $\widehat{\Sigma}_{\text{m},i}$ to obtain $\{ \widehat{y}_{t|t-1} (\widehat{\theta}_i), S_t(\widehat{\theta}_i) \}_{t = 1}^N$ and evaluate the cost function in~\eqref{eq:MLapprox}.
\item \label{algstep:optimization} Determine $\widehat{\theta}_{i+1}$ using the numerical gradient of the cost function in~\eqref{eq:MLapprox}, its approximate Hessian and a backtracking line search algorithm.
\item Obtain $\widehat{D}_{i+1}$, $\widehat{o}_{i+1}$, $\widehat{m}^\text{n}_{i+1}$, $\widehat{\delta}_{\omega,i+1}$, $\widehat{\Sigma}_{\omega,i+1}$, $\widehat{\Sigma}_{\text{a},i+1}$, $\widehat{\Sigma}_{\text{m},i+1}$ from $\widehat{\theta}_{i+1}$. 
\item Set $i := i+1$ and repeat from \Stepref{algstep:ekf} until convergence.
\end{enumerate}
\end{enumerate}
\end{algorithm}

\section{Calibration algorithm}
\label{sec:resultingAlgorithm}
In our magnetometer calibration algorithm we solve the optimization problem~\eqref{eq:MLapprox} to estimate the parameter vector $\theta$. In this section we introduce the resulting calibration algorithm which is summarized in \Algorithmref{alg:ml}. In \Sectionref{sec:optimizationAlg}, we first discuss our optimization strategy. A crucial part of this optimization strategy is the evaluation of the cost function. Some details related to this are discussed in \Sectionref{sec:ekf}. Finally, in \Sectionref{sec:parameterVector} we introduce the parameter vector $\theta$ in more detail.

\subsection{Optimization algorithm}
\label{sec:optimizationAlg}
The optimization problem~\eqref{eq:MLapprox} is solved in \Stepref{algstep:ml} of \Algorithmref{alg:ml}. Standard unconstrained minimization techniques are used, which iteratively update the parameter estimates as
\begin{align}
\theta_{i+1} = \theta_i - \alpha_i \left[ \mathcal{H}(\theta_i) \right]^{-1} \mathcal{G}(\theta_i),
\end{align}
where the \textit{direction} of the parameter update at iteration $i$ is determined by $\left[ \mathcal{H}(\theta_i) \right]^{-1} \mathcal{G}(\theta_i)$. The step \textit{length} of the update at iteration $i$ is denoted by $\alpha_i$. 

Typical choices for the search direction include choosing $\mathcal{G}(\theta_i)$ to be the gradient of the cost function in~\eqref{eq:MLapprox} and $\mathcal{H}(\theta_i)$ to be its Hessian. This leads to a Newton optimization algorithm. However, computing the gradient and Hessian of~\eqref{eq:MLapprox} is not straightforward. Possible approaches are discussed in~\cite{astrom:1980,segalW:1989} for the case of linear models. In the case of nonlinear models, however, they only lead to approximate gradients, see \eg \cite{kokDSW:2015,kokkalaSS:2015}. For this reason we make use of a numerical approximation of $\mathcal{G}(\theta_i)$ instead and use a Broyden-Fletcher-Goldfarb-Shanno (BFGS) method with damped updating \cite{nocedalW:2006} to approximate the Hessian. Hence, the minimization is performed using a quasi-Newton optimization algorithm. A backtracking line search is used to find a good step length $\alpha_i$. 

Proper initialization of the parameters is crucial since the optimization problem~\eqref{eq:MLapprox} is non-convex. \Stepref{algstep:initial} summarizes the three-step process used to obtain good initial estimates of all parameters. 

\subsection{Evaluation of the cost function}
\label{sec:ekf}
An important part of the optimization procedure is the evaluation of the cost function in~\eqref{eq:MLapprox}. This requires running an \ac{ekf} using the state space model~\eqref{eq:greyBoxProblem} to estimate the orientation of the sensor. This \ac{ekf} uses the angular velocity $\omega_t$ as an input to the dynamic model~\eqref{eq:greyBoxProblemDyn}. An estimate of the angular velocity is obtained from the gyroscope measurements $y_{\omega,t}$ which are modeled as
\begin{align}
y_{\omega,t} = \omega_t + \delta_\omega + e_{\omega,t}.
\label{eq:gyrMeasModel}
\end{align}
The measurement model~\eqref{eq:greyBoxProblemMeas} entails the accelerometer measurements and the magnetometer measurements. The magnetometer measurement model can be found in~\eqref{eq:modelMagEq}. The accelerometer measurements $y_{\text{a},t}$ are modeled as
\begin{align}
y_{\text{a},t} = R^{\text{bn}}_t (a_t^\text{n} - g^\text{n}) + e_{\text{a},t} \approx - R^{\text{bn}}_t g^\text{n} + e_{\text{a},t}, 
\label{eq:accMeasModel}
\end{align}
where $a_t^\text{n}$ denotes the sensor's acceleration in the navigation frame and $g^\text{n}$ denotes the earth's gravity. The rotation matrix $R^{\text{bn}}_t$ has previously been introduced in \Sectionref{sec:modeling}.

The state in the \ac{ekf}, which represents the sensor orientation, can be parametrized in different ways. In previous work we have used a quaternion representation as a 4-dimensional state vector~\cite{kokS:2014}. In this work we instead use an implementation of the \ac{ekf}, which is sometimes called a multiplicative \ac{ekf}~\cite{markley:2003,crassidisMC:2007,hol:2011}. Here, a $3$-dimensional state vector represents the \textit{orientation deviation from a linearization point}. More details on this implementation can be found in~\cite{kok:2014}.

The \ac{ekf} returns the one step ahead predicted measurements $\{ \widehat{y}_{t|t-1} (\theta) \}_{t=1}^N$ and their covariance $\{ S_t (\theta) \}_{t=1}^N$ which can be used to evaluate~\eqref{eq:MLapprox}. The cost function needs to be evaluated for the current parameter estimates in \Stepref{algstep:ekf} but also needs to be evaluated once for each component of the parameter vector $\theta$ to compute the numerical gradient. Hence, each iteration $i$ requires running the \ac{ekf} at least $n_\theta+1$ times. Note that the actual number of evaluations can be higher since the backtracking line search algorithm used to determine $\alpha_i$ can require a varying number of additional evaluations. Since $n_\theta = 34$, computing the numerical gradient is computationally rather expensive. However, it is possible to parallelize the computations. 

\subsection{The parameter vector $\theta$}
\label{sec:parameterVector}
As apparent from \Sectionref{sec:modeling}, our main interest lies in determining the calibration matrix $D$ and the offset vector $o$, which can be used to correct the magnetometer measurements to obtain more accurate orientation estimates. To solve the calibration problem, however, we also estimate a number of other parameters. 

First, the local magnetic field $m^\text{n}$ introduced in \Sectionref{sec:modeling} is in general scenarios unknown and needs to be estimated. In outdoor environments, $m^\text{n}$ is equal to the local earth magnetic field and is accurately known from geophysical studies, see \eg \cite{ncei}. In indoor environments, however, the local magnetic field can differ quite significantly from the local earth magnetic field. Because of that, we treat $m^\text{n}$ as an unknown constant. Second, the gyroscope measurements that are used to describe the change in orientation of the sensor in~\eqref{eq:greyBoxProblemDyn} are corrupted by a bias $\delta_\omega$. This bias is slowly time varying but for our relatively short experiments it can be assumed to be constant. Hence, it is treated as part of the parameter vector $\theta$.
Finally, we treat the noise covariance matrices $\Sigma_\omega$, $\Sigma_\text{a}$ and $\Sigma_\text{m}$ as unknown. In summary, the parameter vector $\theta$ consists of
\begin{subequations}
\begin{align}
D &\in \mathbb{R}^{3\times3}, \\
o &\in \mathbb{R}^3, \\
m^\text{n} &\in \{ \mathbb{R}^3 : ||m^\text{n}||_2^2=1, m^\text{n}_x > 0, m^\text{n}_y = 0\}, \label{subeq:constraintmn} \\
\delta_\omega &\in \mathbb{R}^3, \\
\Sigma_\omega &\in \{ \mathbb{R}^{3\times3} : \Sigma_\omega \succeq 0, \Sigma_\omega = \Sigma_\omega^\Transp \}, \label{subeq:constraintSigmaw} \\ 
\Sigma_\text{a} &\in \{ \mathbb{R}^{3\times3} : \Sigma_\text{a} \succeq 0, \Sigma_\text{a} = \Sigma_\text{a}^\Transp \}, \label{subeq:constraintSigmaa} \\
\Sigma_\text{m} &\in \{ \mathbb{R}^{3\times3} : \Sigma_\text{m} \succeq 0, \Sigma_\text{m} = \Sigma_\text{m}^\Transp \}, \label{subeq:constraintSigmam}
\end{align}
\label{eq:constraintsTheta}%
\end{subequations}
where $m_x^\text{n}$ and $m_y^\text{n}$ denote the $x$- and $y$- component of $m^\text{n}$, respectively. The notation $\Sigma \succeq 0$ denotes the assumption that the matrix $\Sigma$ is positive semi-definite.

Although~\eqref{subeq:constraintmn} and \eqref{subeq:constraintSigmaw}~--~\eqref{subeq:constraintSigmam} suggest that constrained optimization is needed, it is possible to circumvent this via suitable reparametrizations. The covariance matrices can be parametrized in terms of their Cholesky factorization, leading to only $6$ parameters for each $3\times3$ covariance matrix. The local magnetic field can be parametrized using only one parameter as in~\eqref{eq:generalEqDipAngle}. Note that in our implementation we prefer to use the representation~\eqref{eq:magfieldDip} for the ML problem~\eqref{eq:MLapprox}. Although this latter parametrization does not account for the constraint $m^\text{n}_x > 0$, this is of no concern due to proper initialization. The procedure to obtain good initial estimates of all parameters is the topic of the next section. 
 
\section{Finding good initial estimates}
\label{sec:initialEstimate}
Since the optimization problem is non-convex, the parameter vector $\theta$ introduced in \Sectionref{sec:resultingAlgorithm} needs proper initialization. An initial estimate $\widehat{\theta}_0$ is obtained using a three-step method. As a first step, the gyroscope bias~$\delta_\omega$ and the noise covariances of the inertial sensors, $\Sigma_\omega$, $\Sigma_\text{a}$, and of the magnetometer, $\Sigma_\text{m}$, are initialized. This is done using a short batch of stationary data. Alternatively, they can be initialized based on prior sensor knowledge. As a second step, described in \Sectionref{sec:initEllipse}, an ellipsoid fitting problem is solved using the magnetometer data. This maps the ellipsoid of data to a sphere but can not determine the rotation of the sphere. The rotation of the sphere is determined in a third step of the initialization procedure. This step also determines an initial estimate of the normalized local magnetic field $m^\text{n}$. 

\subsection{Ellipsoid fitting}
\label{sec:initEllipse}
Using the definition of the normalized local magnetic field $m^\text{n}$, we would expect all calibrated magnetometer measurements to lie on the unit sphere,  
\begin{align}
\| m^\text{n} \|_2^2 - 1 &= \| R^{\text{bn}}_t m^\text{n} \|_2^2 - 1 \nonumber \\ 
&= \| D^{-1} \left( y_{\text{m},t} - o  - e_{\text{m},t} \right) \|_2^2 - 1 = 0.
\label{eq:Do_ellipseFit}
\end{align}
In practice, the measurements are corrupted by noise and the equality~\eqref{eq:Do_ellipseFit} does not hold exactly. The ellipsoid fitting problem can therefore be written as
\begin{align}
y_{\text{m},t}^\Transp A  y_{\text{m},t}  + b^\Transp  y_{\text{m},t}  + c \approx 0,
\label{eq:abc_ellipseFit}
\end{align}
with
\begin{subequations}
\begin{align}
A &\triangleq D^{-\Transp} D^{-1}, \\
b &\triangleq - 2 o^\Transp D^{-\Transp} D^{-1}, \\
c &\triangleq o^\Transp D^{-\Transp} D^{-1} o. 
\label{eq:defAbc}
\end{align}
\end{subequations}
Assuming that the matrix $A$ is positive definite, this can be recognized as the definition of an ellipsoid with parameters $A$, $b$ and $c$ (see \eg \cite{ganderGS:1994}). We can rewrite~\eqref{eq:abc_ellipseFit} as a linear relation of the parameters as
\begin{align}
M \xi \approx 0, 
\end{align}
with
\begin{align}
M = \begin{pmatrix} y_{\text{m},1} \otimes y_{\text{m},1} & y_{\text{m},1} & 1 \\ y_{\text{m},2} \otimes y_{\text{m},2} & y_{\text{m},2} & 1 \\ \vdots & \vdots & \vdots  \\ y_{\text{m},N} \otimes y_{\text{m},N} & y_{\text{m},N} & 1 \end{pmatrix}, \quad \xi = \begin{pmatrix} \vect{A} \\ b \\ c \end{pmatrix},
\end{align}
where $\otimes$ denotes the Kronecker product and $\vect$ denotes the vectorization operator. This problem has infinitely many solutions and without constraining the length of the vector $\xi$, the trivial solution $\xi = 0$ would be obtained. A possible approach to solve the ellipsoid fitting problem is to make use of a singular value decomposition~\cite{ganderGS:1994,kokHSGL:2012}. This approach inherently poses a length constraint on the vector $\xi$, assuming that its norm is equal to $1$. It does, however, not guarantee positive definiteness of the matrix $A$. Although positive definiteness of $A$ is not guaranteed, there are only very few practical scenarios in which the estimated matrix $A$ will not be positive definite. A non-positive definite matrix $A$ can for instance be obtained in cases of very limited rotation of the sensor. The problem of allowing a non-positive definite matrix $A$ can be circumvented by instead solving the ellipsoid fitting problem as a semidefinite program~\cite{calafiore:2002,boydV:2004} 
\begin{equation}
\begin{aligned}
\min_{A,b,c} \quad & \tfrac{1}{2} \| M \begin{pmatrix} \vect A \\ b \\ c \end{pmatrix} \|_2^2, \\ 
\text{s.t.} \quad & \Tr{A} = 1, \quad  A \in S_{++}^{3 \times 3}, 
\label{eq:ellipseFittingOptProblem}
\end{aligned}
\end{equation} 
where $S^{3 \times 3}_{++}$ denotes the set of $3 \times 3$ positive definite symmetric matrices. By constraining the trace of the matrix $A$, \eqref{eq:ellipseFittingOptProblem} avoids the trivial solution of $\xi = 0$. The problem~\eqref{eq:ellipseFittingOptProblem} is a convex optimization problem and therefore has a globally optimal solution and does not require an accurate initial guess of the parameter vector~$\xi$. The optimization problem can easily be formulated and efficiently solved using freely available software packages like YALMIP~\cite{lofberg:2004} or CVX~\cite{cvx}. 

Initial estimates of the calibration matrix $D$ and the offset vector $o$ can be obtained from the estimated $\widehat{A},\widehat{b},\widehat{c}$ as
\begin{subequations}
\begin{align}
\beta &= \left( \tfrac{1}{4} \widehat{b}^\Transp \widehat{A}^{-1} \widehat{b} - \widehat{c} \right)^{-1}, \\
\widetilde{D}_0^\Transp \widetilde{D}_0 &= \beta \widehat{A}^{-1}, \label{eq:initialDtilde}\\
\widehat{o}_0 &= \tfrac{1}{2} \widehat{A}^{-1} \widehat{b}, \label{eq:initialo}
\end{align}
\end{subequations}
where $\widehat{o}_0$ denotes the initial estimate of the offset vector $o$. From~\eqref{eq:initialDtilde} it is not possible to uniquely determine the initial estimate of the calibration matrix $D$. We determine an initial estimate of the calibration matrix $D$ using a
Cholesky decomposition, leading to a lower triangular $\widetilde{D}_0$. However, any $\widetilde{D}_0 U$ where $UU^\Transp = \mathcal{I}_3$ will also fulfill~\eqref{eq:initialDtilde}. As discussed in \Assumptionref{ass:Do} in \Sectionref{sec:modeling}, we assume that the sensor axes of the inertial sensors and the magnetometers are related by a rotation, implying that we restrict the matrix $U$ to be a rotation matrix. The initial estimate $\widehat{D}_0$ can therefore be defined in terms of $\widetilde{D}_0$ as
\begin{align}
\widehat{D}_0 = \widetilde{D}_0 R_\text{D}.
\label{eq:D0tildeRot}
\end{align}
The unknown rotation matrix $R_\text{D}$ will be determined in \Sectionref{sec:initMisalignment}. 

\subsection{Determine misalignment of the inertial and magnetometer sensor axes}
\label{sec:initMisalignment}
The third step of the initial estimation aims at determining the misalignment between the inertial and the magnetometer sensor axes. It also determines an initial estimate of the normalized local magnetic field $\widehat{m}_0^\text{n}$. These estimates are obtained by combining the magnetometer measurements with the inertial sensor measurements. The approach is based on the fact that the inner product of two vectors is invariant under rotation. The two vectors considered here are $m^\text{n}$ and the vertical $v^\text{n} = \begin{pmatrix} 0 & 0 & 1 \end{pmatrix}^\Transp$. Hence, it is assumed that the inner product of the vertical $v^\text{b}_t$ in the body frame $b$, 
\begin{subequations}
\begin{align}
v^\text{b}_t &= R^{\text{bn}}_t v^{\text{n}},
\label{subeq:vertical}
\intertext{and the normalized local magnetic field $m^\text{b}_t$ in the body frame,}
m^\text{b}_t &= R_\text{D}^\Transp \widetilde{D}_0^{-1} \left( y_{\text{m},t} - \widehat{o}_0 \right),
\label{subeq:malign}
\end{align}
\end{subequations}
is constant. The matrix $R_\text{D}$ in~\eqref{subeq:malign} denotes the rotation needed to align the inertial and magnetometer sensor axes. The rotation matrix $R^{\text{nb}}_t$ in~\eqref{subeq:vertical} is a rotation matrix representation of the orientation estimate at time $t$ obtained from an \ac{ekf}. This \ac{ekf} is similar to the one described in \Sectionref{sec:ekf}. It does not use the magnetometer measurements, since they have not properly been calibrated yet and can therefore not result in accurate heading estimates. However, to determine the vertical $v^\text{b}_t$, only the sensor's inclination is of concern, which can be determined using the inertial measurements only. 

The inner product between $m^\text{n}$ and $v^\text{n}$ is equal to $m^\text{n}_z$ (see also~\eqref{eq:magfieldVertical}). Since this inner product is invariant under rotation, we can formulate the following minimization problem
\begin{align}
\label{eq:misalignmentDetermination}
\min_{R_\text{D},m^\text{n}_{z,0}}& \quad && \tfrac{1}{2} \sum_{t=1}^N \| m^\text{n}_{z,0} - \left(v^{\text{n}}\right)^\Transp R^{\text{nb}}_t R_\text{D}^\Transp \widetilde{D}_0^{-1} \left( y_{\text{m},t} - \widehat{o}_0 \right) \|_2^2, \nonumber \\ 
\text{s.t.}& \quad && R_\text{D} \in \SO{3}.
\end{align} 
The rotation matrix $R_\text{D}$ can be parametrized using an orientation deviation from a linearization point similar to the approach described in \Sectionref{sec:ekf}. Hence,~\eqref{eq:misalignmentDetermination} can be solved as an unconstrained optimization problem. 

Based on these results and~\eqref{eq:D0tildeRot} we obtain the following initial estimates 
\begin{subequations}
\begin{align}
\widehat{D}_0 &= \widetilde{D}_0 \widehat{R}_\text{D}, \label{eq:initialD} \\
\widehat{m}_0^\text{n} &= \begin{pmatrix} \sqrt{1-\left(\widehat{m}^\text{n}_{z,0}\right)^2} & 0 & \widehat{m}^\text{n}_{z,0} \end{pmatrix}^\Transp. \label{eq:initialmn}
\end{align}
\end{subequations}
Hence, we have obtained an initial estimate $\widehat{\theta}_0$ of the entire parameter vector $\theta$ as introduced in \Sectionref{sec:resultingAlgorithm}. 

\section{Experimental results} 
\subsection{Experimental setup}
\begin{figure}
\begin{center}
\includegraphics[width = 0.48\columnwidth]{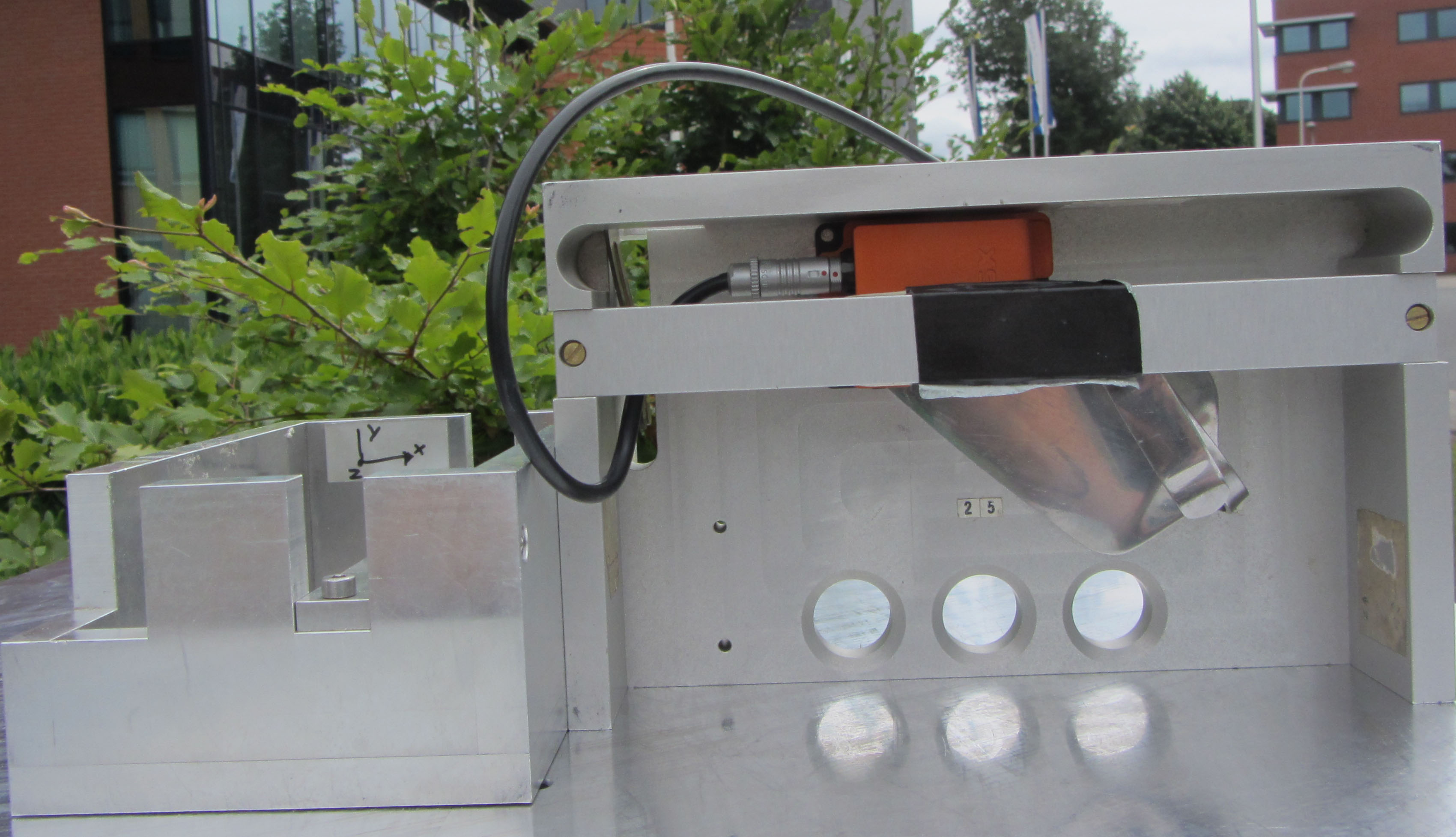} \\ \vspace{1mm}
\includegraphics[width = 0.48\columnwidth]{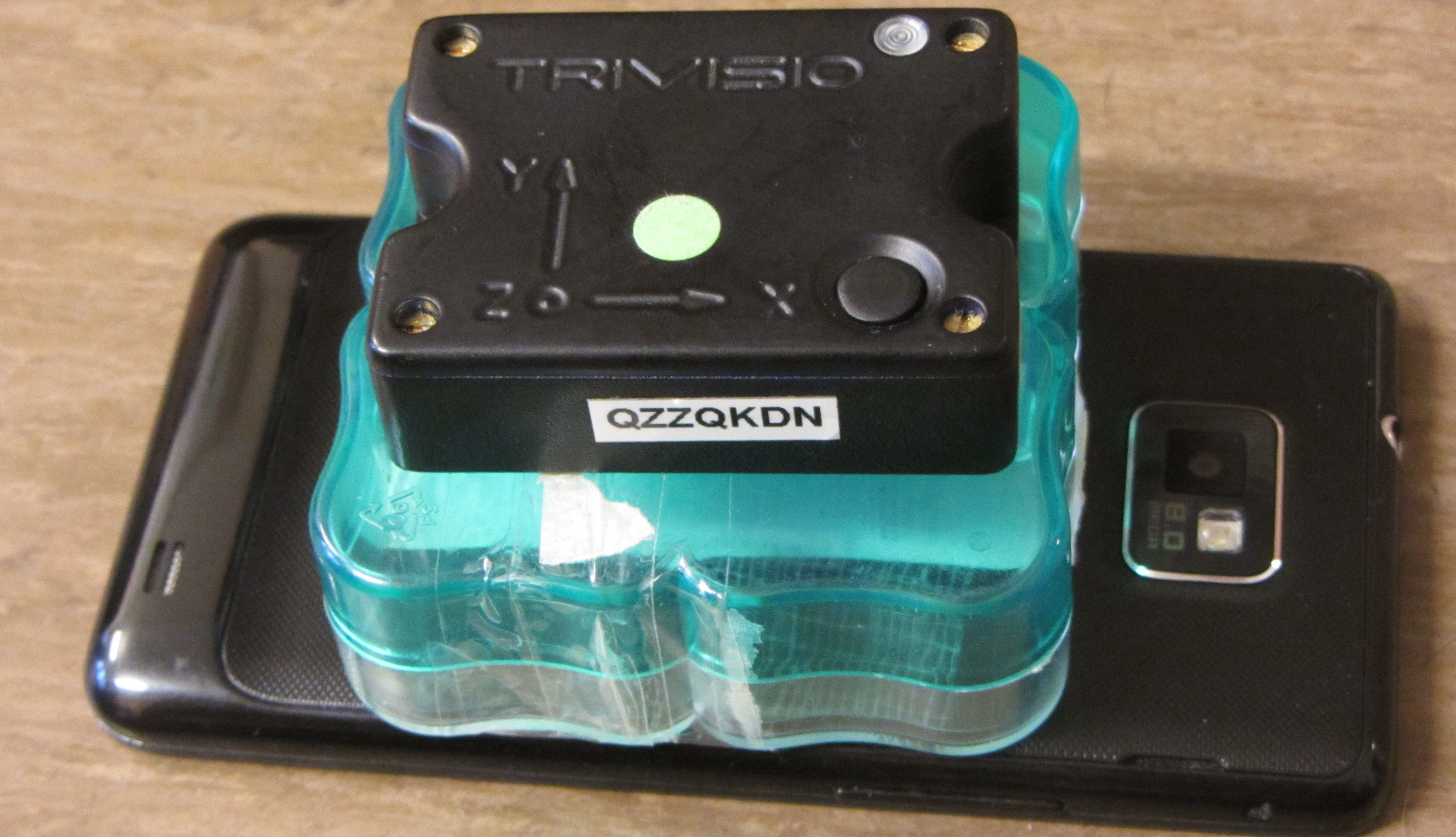}%
\caption{Top: experimental setup where a calibration experiment is performed outdoors. An Xsens MTi-100 IMU (orange box) together with a magnetic disturbance is placed in an aluminum block. Bottom: experimental setup using a Trivisio Colibri Wireless IMU (black box). A phone is used as a source of magnetic disturbance. To avoid saturation of the magnetometer, the phone is not attached directly to the IMU.}  
\label{fig:experimentalSetups}                                 
\end{center}                             
\end{figure}

Experiments have been performed using two commercially available inertial measurements units (IMUs), an Xsens~\mbox{MTi-100}~\cite{xsens} and a Trivisio Colibri Wireless IMU~\cite{trivisio}. The experimental setup of both experiments can be found in \Figureref{fig:experimentalSetups}. The experiment with the Xsens IMU was performed outdoors to ensure a homogeneous local magnetic field. The experiment with the Trivisio IMU was performed indoors. However, the experiment was performed relatively far away from any magnetic materials such that the local magnetic field is as homogenous as possible. The Xsens IMU was placed in an aluminum block with right angles which can be used to rotate the sensor $90^\circ$ to verify the heading results. For both sensors, inertial and magnetometer measurements were collected at $100$\,Hz.  

\subsection{Calibration results}
\begin{figure}
	\centering
	\includegraphics[width = 0.4\columnwidth]{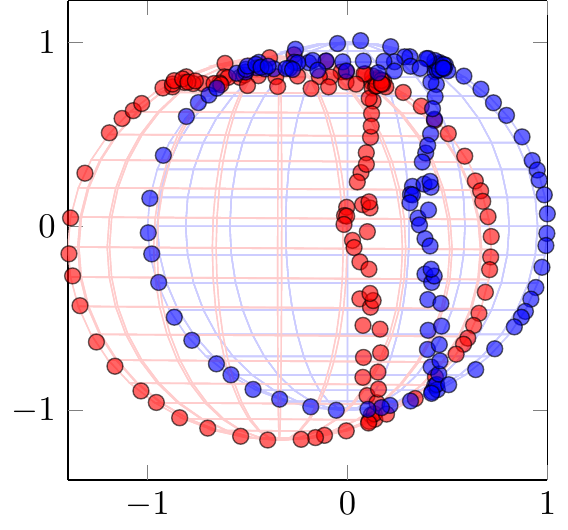}
	\caption{Calibration results from the experiment with the Trivisio IMU. The ellipsoid of magnetometer data (red) lies on a unit sphere after calibration (blue).}
	\label{fig:ellipseTrivisio}
\end{figure}

\begin{figure}
	\begin{center}
	\includegraphics[width = 0.48\columnwidth]{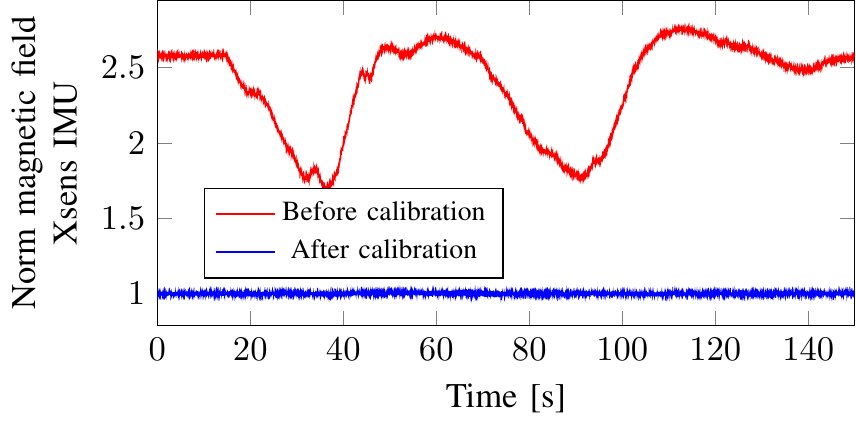}
	\includegraphics[width = 0.48\columnwidth]{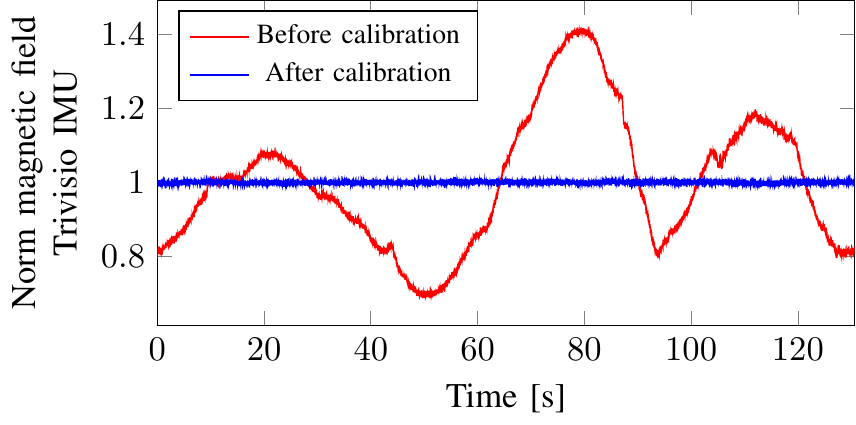}
	\caption{Norm of the magnetic field measurements before (red) and after (blue) calibration for (top) the experiment with the Xsens IMU and for (bottom) the experiment with the Trivisio IMU.}
	\label{fig:normMagField}
	\end{center}
\end{figure}

For calibration, the IMU needs to be slowly rotated such that the assumption of zero acceleration is reasonably valid. This leads to an ellipsoid of magnetometer data as depicted in red in Figs.~\ref{fig:ellipseXsens} and~\ref{fig:ellipseTrivisio}. Note that for plotting purposes the data has been downsampled to $1$\,Hz. To emphasize the deviation of the norm from 1, the norm of the magnetometer data is depicted in red in \Figureref{fig:normMagField} for both experiments.

For the experiment with the Xsens IMU, the following calibration matrix $\widehat{D}$ and offset vector $\widehat{o}$ are found
\begin{align}
\label{eq:mlEstimatesXsens}
\widehat{D} = \begin{pmatrix} 0.74 &  -0.13 & 0.01 \\
   -0.12 & 0.68  & 0.01 \\
   -0.03 & 0.43 & 1.00 
   \end{pmatrix}, \quad \widehat{o} = \begin{pmatrix}  1.36 \\
    1.22 \\
   -0.94 \end{pmatrix}   
\end{align}
using \Algorithmref{alg:ml}. Applying the calibration result to the magnetometer data leads to the unit sphere of data in blue in \Figureref{fig:ellipseXsens}. The norm of the magnetometer data after calibration can indeed be seen to lie around 1, as depicted in blue in \Figureref{fig:normMagField}. 

\begin{figure}
	\centering
	\captionsetup[subfigure]{oneside}
	\subfloat[Xsens IMU, estimation data]{%
		\includegraphics[width = 0.3\columnwidth]{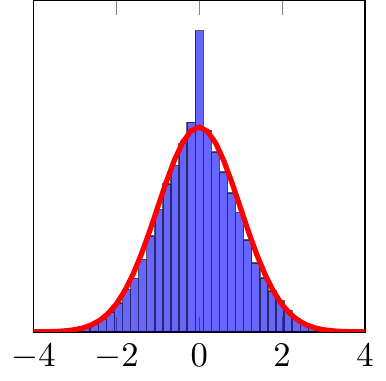}
		\label{fig:normResXsens-est}
        } 
        	\subfloat[Xsens IMU, validation data]{%
		\includegraphics[width = 0.3\columnwidth]{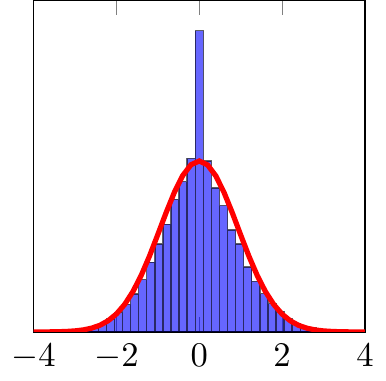}
		\label{fig:normResXsens-val}
        } 
	\caption{Histogram of the normalized residuals $S_{t}^{-1/2} (y_t - \widehat{y}_{t|t-1})$ from the \ac{ekf} after calibration for the estimation data set (left) and for a validation data set (right) for the experiments performed with the Xsens IMU. A Gaussian distribution (red) is fitted to the data.}
	\label{fig:normResXsens}
\end{figure}

As a measure of the calibration quality, we analyze the normalized residuals $S_{t}^{-1/2} (y_t - \widehat{y}_{t|t-1})$ after calibration from the \ac{ekf}. For each time $t$, this is a vector in $\mathbb{R}^6$. In the case of correctly calibrated parameters that sufficiently model the magnetic disturbances, we expect the stacked normalized residuals $\{ S_{t}^{-1/2} (y_t - \widehat{y}_{t|t-1}) \}_{t=1}^N \in \mathbb{R}^{6N}$ to be normally distributed with zero mean and standard deviation~1. The histogram and a fitted Gaussian distribution can be found in \Figureref{fig:normResXsens-est}. The residuals resemble a $\mathcal{N}(0,1)$ distribution except for the large peak around zero and -- not visible in the plot -- a small amount of outliers outside of the plotting interval. This small amount of outliers is due to the fact that there are a few measurement outliers in the accelerometer data. Large accelerations can for instance be measured when the setup is accidentally bumped into something and violate our assumption that the acceleration of the sensor is approximately zero. We believe that the peak around zero is due to the fact that the algorithm compensates for the presence of the large residuals.

To analyze if the calibration is also valid for a different (validation) data set with the same experimental setup, the calibrated parameters have been used on a second data set. Figures of the ellipsoid of magnetometer data and the sphere of calibrated magnetometer data are not included since they look very similar to Figs.~\ref{fig:ellipseXsens} and~\ref{fig:normMagField}. The residuals after calibration of this validation data set can be found in \Figureref{fig:normResXsens-val}. The fact that these residuals look very similar to the ones for the original data suggests that the calibration parameters obtained are also valid for this validation data set.

\begin{figure}
	\centering
	\captionsetup[subfigure]{oneside}
	\subfloat[Trivisio IMU, estimation data]{%
		\includegraphics[width = 0.3\columnwidth]{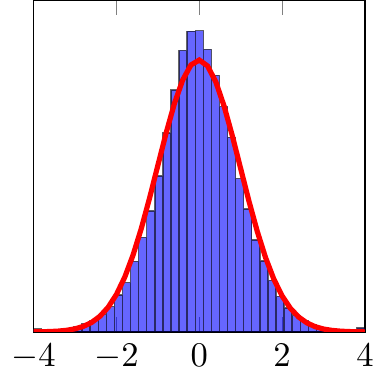}
		\label{fig:normResTrivisio-est}
        } 
        	\subfloat[Trivisio IMU, validation data]{%
		\includegraphics[width = 0.3\columnwidth]{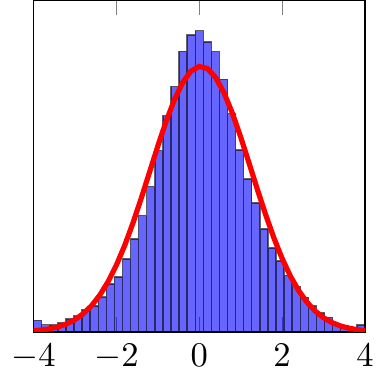}
		\label{fig:normResTrivisio-val}
        } 
	\caption{Histogram of the normalized residuals $S_{t}^{-1/2} (y_t - \widehat{y}_{t|t-1})$ from the \ac{ekf} after calibration for the estimation data set (left) and for a validation data set (right) for the experiments performed with the Trivisio IMU. A Gaussian distribution (red) is fitted to the data.}
	\label{fig:normResTrivisio}
\end{figure}

The Trivisio IMU outputs the magnetometer data in microtesla. Since our algorithm scales the calibrated measurements to a unit norm, the obtained $\widehat{D}$ and offset vector $\widehat{o}$ from \Algorithmref{alg:ml} are in this case of much larger magnitude, 
\begin{align}
\widehat{D} = \begin{pmatrix}  61.74 & 0.59 & 0.09 \\
   -1.01 & 60.74 & 0.23 \\
   -0.39 & 0.06 & 60.80
   \end{pmatrix}, \quad \widehat{o} = \begin{pmatrix}    -19.77 \\
   -1.68 \\
   -6.98 \end{pmatrix}.   
\end{align}
The sphere of calibrated data and its norm can be found in blue in Figs.~\ref{fig:ellipseTrivisio} and~\ref{fig:normMagField}. Note that for plotting purposes, the magnetometer data before calibration is scaled such that its mean lies around 1. The obtained $\widehat{D}$ and $\widehat{o}$ are scaled accordingly to plot the red ellipsoid in \Figureref{fig:ellipseTrivisio}. The normalized residuals $S_{t}^{-1/2} (y_t - \widehat{y}_{t|t-1})$ of the \ac{ekf} using both the estimation and a validation data set are depicted in \Figureref{fig:normResTrivisio}. For this data set, the accelerometer data does not contain any outliers and the residuals resemble a $\mathcal{N}(0,1)$ distribution fairly well.

From these results we can conclude that \Algorithmref{alg:ml} gives good magnetometer calibration results for experimental data from two different commercially available IMUs. A good fit of the ellipsoid of data to a sphere is obtained and the algorithm seems to give good estimates analyzed in terms of its normalized residuals. Since magnetometer calibration is generally done to obtain improved heading estimates, it is important to also interpret the quality of the calibration in terms of the resulting heading estimates. In \Sectionref{sec:exp_headingVal} this will be done based on experimental results. The heading performance will also be analyzed based on simulations in \Sectionref{sec:simulationResults}.  
 
\subsection{Heading estimation}
\label{sec:exp_headingVal}
\begin{figure}
	\centering
	\includegraphics[width = 0.6\columnwidth]{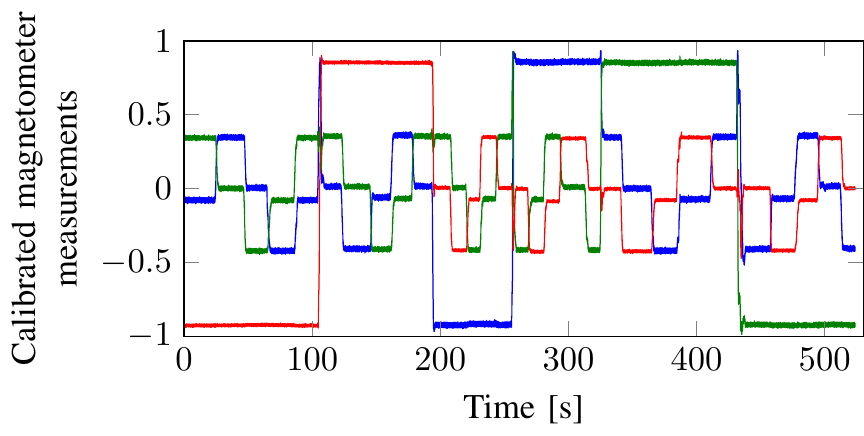}
	\caption{Calibrated magnetometer data of an experiment rotating the sensor into $24$ different sensor orientations where the blue, green and red lines represent the data from the $x$-, $y$- and $z$-axis of the magnetometer, respectively.}
	\label{fig:magCalRotations}
\end{figure}

An important goal of magnetometer calibration is to facilitate good heading estimates. To check the quality of the heading estimates after calibration, the block in which the Xsens IMU was placed (shown in \Figureref{fig:experimentalSetups}) is rotated around all axes. This block has right angles and it can therefore be placed in $24$ orientations that differ from each other by $90$ degrees. The experiment was conducted in Enschede, the Netherlands. The dip angle $\delta$ at this location is approximately $67^\circ$~\cite{ncei}. Hence, we expect the calibrated magnetometer measurements to resemble rotations of the normalized magnetic field $m^\text{n} = \begin{pmatrix} 0.39 & 0 & -0.92 \end{pmatrix}^\Transp$ (see also~\eqref{eq:rotationEarthMagField} and~\eqref{eq:magfieldDip}). The calibrated magnetometer data from the experiment is shown in \Figureref{fig:magCalRotations} and consists of the following stationary time periods:
\begin{description}[style=unboxed]
\item[$\boldsymbol{z}$-axis up] During the period $0 - 105$s, the magnetometer is flat with its $z-$axis pointing upwards. Hence, the $z$-axis (red) of the magnetometer measures the vertical component of the local magnetic field $m^\text{n}_z$. During this period, the sensor is rotated by $90^\circ$ around the $z$-axis into $4$ different orientations and subsequently back to its initial orientation. This results in the 5 steps for measurements in the $x$- (blue) and $y$-axis (green) of the magnetometer. 
\item[$\boldsymbol{z}$-axis down] A similar rotation sequence is performed with the block upside down at $110 - 195$s, resulting in a similar pattern for measurements in the $x$- and $y$-axis of the magnetometer. During this time period, the $z$-axis of the magnetometer measures $-m^\text{n}_z$ instead. 
\item[$\boldsymbol{x}$-axis up] The procedure is repeated with the $x$-axis of the sensor pointing upwards during the period $200-255$s, rotating around the $x$-axis into 4 different orientations and back to the initial position. This results in the 5 steps for measurements in the $y$- and $z$-axis of the magnetometer. 
\item[$\boldsymbol{x}$-axis down] A similar rotation sequence is performed with the $x$-axis pointing downwards at $265 - 325$ seconds.
\item[$\boldsymbol{y}$-axis down] Placing the sensor with the $y$-axis downwards and rotating around the $y$-axis results in the data at $350 - 430$ seconds. The rotation results in the 5 steps for measurements in the $x$- and $z$-axis of the magnetometer. 
\item[$\boldsymbol{y}$-axis up] A similar rotation sequence is performed with the $y$-axis pointing upwards at $460 - 520$ seconds.
\end{description}
Since the experimental setup was not placed \emph{exactly} vertical, it is not possible to compare the \emph{absolute orientations}. However, it is possible to compare the \emph{difference in orientation} which is known to be $90^\circ$ due to the properties of the block in which the sensor was placed. To exclude the effect of measurement noise, for each of the stationary periods in \Figureref{fig:magCalRotations}, 500 samples of magnetometer and accelerometer data are selected. Their mean values are used to estimate the orientation of the sensor. Here, the accelerometer data is used to estimate the inclination. The heading is estimated from the horizontal component of the magnetometer data. This procedure makes use of the fact that the orientation of the sensor can be determined from two linearly independent vectors in the navigation frame -- the gravity and the direction of the magnetic north -- and in the body frame -- the mean accelerometer and magnetometer data. It is referred to as the TRIAD algorithm~\cite{shusterO:1981}. \Tableref{table:rotations} reports the deviation from $90^\circ$ between two subsequent rotations. Note that the metal object causing the magnetic disturbance as shown in \Figureref{fig:experimentalSetups} physically prevents the setup from being properly placed in all orientations around the $y$-axis. Rotation around the $y$-axis with the $y$-axis pointing upwards has therefore not been included in \Tableref{table:rotations}. 

\tikzsetnextfilename{magCalRotations}

\begin{table*}[t]
\caption{Difference in estimated heading between two subsequent rotations around the sensor axes using calibrated magnetometer data. The values represent the deviation in degrees from $90^\circ$. Included are both the results using the \ac{ml} estimates from Algorithm~\ref{alg:ml} and the results using initial estimates from \Stepref{algstep:initial} in the algorithm.}
\begin{center}
\begin{tabular}{cccccccccc}
\toprule
\multicolumn{4}{c}{$z$-axis} & \multicolumn{4}{c}{$x$-axis} & \multicolumn{2}{c}{$y$-axis} \\
\midrule
\multicolumn{2}{c}{$z$ up} & \multicolumn{2}{c}{$z$ down} & \multicolumn{2}{c}{$x$ up} & \multicolumn{2}{c}{$x$ down} & \multicolumn{2}{c}{$y$ down} \\
\cmidrule(lr){1-2}
\cmidrule(lr){3-4}
\cmidrule(lr){5-6}
\cmidrule(lr){7-8}
\cmidrule(lr){9-10}
ML & init & ML & init & ML & init  & ML & init  & ML & init 
\\ \midrule
0.11 & 0.36 & 0.69 & 1.34 & 0.22 & 0.16 & 0.86 & 1.01 & 0.18 & 1.57 
\\
0.22 & 0.90 & 2.48 & 4.36 & 0.07 & 0.20 & 1.57 & 1.45 & 0.29 & 0.76 
\\
0.46 & 1.52 & 1.53 & 3.57 & 0.97 & 0.94 & 0.61 & 0.71 & 0.20 & 0.78 
\\
0.30 & 0.94 & 1.92 & 2.40 & 0.29 & 0.59 & 1.78 & 1.70 & 0.50 & 0.45 
\\
\bottomrule
\end{tabular}
\end{center}
\label{table:rotations}
\end{table*}

Our experiment investigates both the heading errors and the improvement of the heading estimates over the ones obtained after the initial calibration, \ie \Stepref{algstep:initial} in \Algorithmref{alg:ml}. In \Tableref{table:rotations} we therefore include both the heading errors using the initial parameter estimates $\widehat{D}_0$~\eqref{eq:initialD} and $\widehat{o}_0$~\eqref{eq:initialo} and the heading errors using \ac{ml} parameter estimates $\widehat{D}$ and $\widehat{o}$~\eqref{eq:mlEstimatesXsens} obtained using \Algorithmref{alg:ml}. As can be seen, the deviation from $90^\circ$ is small, indicating that good heading estimates are obtained after calibration. Also, the heading estimates using the initial parameter estimates are already fairly good. The mean error is reduced from $1.28^\circ$ for the initial estimate to $0.76^\circ$ for the \ac{ml} estimate. The maximum error is reduced from $4.36^\circ$ for the initial estimate to $2.48^\circ$ for the \ac{ml} estimate. Note that the results of the \ac{ml} estimate from \Algorithmref{alg:ml} are slightly better than the results previously reported by~\cite{kokS:2014}. This can be attributed to the fact that we now use orientation error states instead of the quaternion states in the \ac{ekf} (see \Sectionref{sec:ekf}). This results in slightly better estimates, but also in a smoother convergence of the optimization problem. The quality of the heading estimates is studied further in \Sectionref{sec:simulationResults} based on a simulation study.

\section{Simulated heading accuracy}
\label{sec:simulationResults}
Magnetometer calibration is typically performed to improve the heading estimates. It is, however, difficult to check the heading accuracy experimentally. In~\Sectionref{sec:exp_headingVal}, for instance, we are limited to doing the heading validation on a different data set and we have a limited number of available data points. To get more insight into the orientation accuracy that is gained by executing all of \Algorithmref{alg:ml}, compared to just its initialization phase (\Stepref{algstep:initial} in the algorithm), we engage in a simulation study. In this study we focus on the \ac{rms} heading error for different simulated sensor qualities (in terms of the noise covariances and the gyroscope bias) and different magnetic field disturbances (in terms of different values for the calibration matrix $D$ and offset vector~$o$).

In our simulation study, we assume that the local magnetic field is equal to that in Link\"oping, Sweden. The calibration matrix $D$, the offset vector $o$ and the sensor properties in terms of the gyroscope bias and noise covariances are all sampled from a uniform distribution. The parameters of the distributions from which the sensor properties are sampled are chosen as physically reasonable values as considered from the authors' experience. The noise covariance matrices $\Sigma_\omega$, $\Sigma_\text{a}$ and $\Sigma_\text{m}$ are assumed to be diagonal with three different values on the diagonal. The calibration matrix $D$ is assumed to consist of three parts, 
\begin{align}
D &= D_\text{diag} D_\text{skew} D_\text{rot}, 
\end{align}
where $D_\text{diag}$ is a diagonal matrix with elements $D_{11}, D_{22}, D_{33}$ and $D_\text{rot}$ is a rotation matrix around the angles $\psi, \theta, \phi$. The matrix $D_{\text{skew}}$ models the non-orthogonality of the magnetometer axes as
\begin{align}
D_\text{skew} = \begin{pmatrix} 1 & 0 & 0 \\ \sin \zeta & \cos \zeta & 0 \\ -\sin \eta & \cos \eta \sin \rho & \cos \eta \cos \rho \end{pmatrix},
\end{align}
where the angles $\zeta$, $\eta$, $\rho$ represent the different non-orthogonality angles. The exact simulation conditions are summarized in \Tableref{table:simConditions}.

\begin{table*}[t]
\caption{Settings used in the Monte Carlo simulation.}
\begin{center}
\begin{tabular}{cccc}
\toprule
$D_\text{diag}$ & $D_\text{skew}$ & $D_\text{rot}$ & $o$ \\
\midrule
$D_{11}, D_{22}, D_{33}$ & $\zeta, \eta, \rho$ & $\psi, \theta, \phi$  & $o_1, o_2, o_3$ \\
$\sim \mathcal{U}(0.5,1.5)$ & $\sim \mathcal{U}(-30^\circ,30^\circ)$ & $\sim \mathcal{U}(-10^\circ,10^\circ)$ & $\sim \mathcal{U}(-1,1)$ 
\\
\bottomrule
\vspace{3mm}
\end{tabular}
\begin{tabular}{cccc}
\toprule
$\delta_\omega$ & $\Sigma_\omega$ & $\Sigma_\text{a}$ & $\Sigma_\text{m}$ \\
\midrule
$\delta_{\omega,1}, \delta_{\omega,2}, \delta_{\omega,3}$ & $\Sigma_{\omega,1}, \Sigma_{\omega,2}, \Sigma_{\omega,3}$ & $\Sigma_{\text{a},1}, \Sigma_{\text{a},2}, \Sigma_{\text{a},3}$ & $\Sigma_{\text{m},1}, \Sigma_{\text{m},2}, \Sigma_{\text{m},3}$ \\
$\sim \mathcal{U}(-1,1)$ & $\sim \mathcal{U}(10^{-3},10^{-2})$ & $\sim \mathcal{U}(10^{-3},10^{-1})$ & $\sim \mathcal{U}(10^{-3},10^{-1})$ 
\\
\bottomrule
\end{tabular}
\end{center}
\label{table:simConditions}
\end{table*}

The simulated data consists of $100$ samples of stationary data and subsequently $300$ samples for rotation around all three axes. It is assumed that the rotation is exactly around the origin of the accelerometer triad, resulting in zero acceleration during the rotation. The first $100$ samples are used to obtain an initial estimate of the gyroscope bias $\widehat{\delta}_{\omega,0}$ by computing the mean of the stationary gyroscope samples. The covariance matrices $\widehat{\Sigma}_{\omega,0}$, $\widehat{\Sigma}_{\text{a},0}$ and $\widehat{\Sigma}_{\text{m},0}$ are initialized based on the covariance of these first $100$ samples. The initial estimate then consists of these initial estimates $\widehat{\delta}_{\omega,0}$, $\widehat{\Sigma}_{\omega,0}$, $\widehat{\Sigma}_{\text{a},0}$, $\widehat{\Sigma}_{\text{m},0}$ and the initial calibration matrix $\widehat{D}_0$~\eqref{eq:initialD}, the initial offset vector $\widehat{o}_0$~\eqref{eq:initialo} and the initial estimate of the local magnetic field $m^\text{n}_0$~\eqref{eq:initialmn}. 

To study the heading accuracy, the \ac{ekf} as described in \Sectionref{sec:ekf} is run with both the initial parameter values $\widehat{\theta}_0$ and their \ac{ml} values $\widehat{\theta}^{\text{ML}}$. The orientation errors $\Delta q_t$, encoded as a unit quaternion are computed using 
\begin{align}
\Delta q_t = \widehat{q}^\text{nb}_t \odot \left( q^\text{nb}_{\text{ref},t} \right)^c, 
\end{align}
where $\odot$ denotes a quaternion multiplication and the superscript $c$ denotes the quaternion conjugate (see \eg \cite{hol:2011}). It is computed from the orientation $\widehat{q}^\text{nb}_t$ estimated by the \ac{ekf} and the ground truth orientation $q^\text{nb}_{\text{ref},t}$. Computing the orientation errors in this way is equivalent to subtracting Euler angles in the case of small angles. However, it avoids subtraction problems due to ambiguities in the Euler angles representation. To interpret the orientation errors $\Delta q_t$, they are converted to Euler angles. We focus our analysis on the heading error, \ie on the third component of the Euler angles. 

\begin{figure}
	\centering
	\captionsetup[subfigure]{oneside}
	\subfloat[Initial parameter estimate]{%
		\includegraphics[width = 0.3\columnwidth]{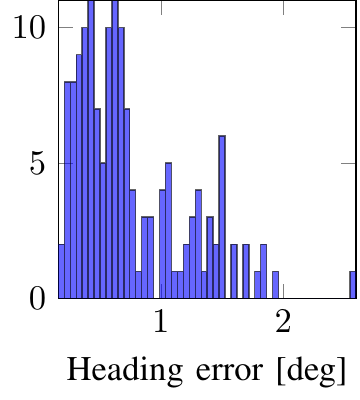}
        } 
        	\subfloat[ML parameter estimate]{%
		\includegraphics[width = 0.3\columnwidth]{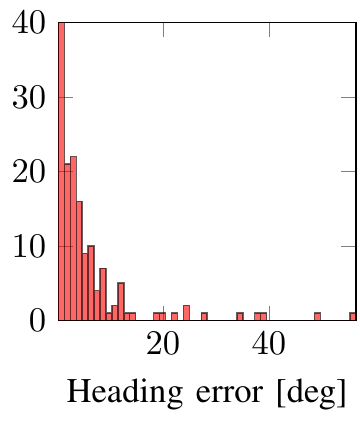}
        } 
	\caption{Histogram of the heading RMSE using the \ac{ml} parameter estimate from Algorithm~\ref{alg:ml} (left, blue) and the initial parameter estimate from Step~\ref{algstep:initial} in the algorithm (right, red). Note the different scales in the two plots.}
	\label{fig:simulatedHeadingRMSE}
\end{figure}

The \ac{rms} of the heading error is plotted for $150$ Monte Carlo simulations in \Figureref{fig:simulatedHeadingRMSE}. As can be seen, the heading \ac{rmse} using the estimate of the calibration parameters from \Algorithmref{alg:ml} is consistently small. The heading \ac{rmse} based on the initialization phase in \Stepref{algstep:initial} of the algorithm, however, has a significantly larger spread. This clearly shows that orientation accuracy can be gained by executing all of \Algorithmref{alg:ml}. Note that in all simulations, analysis of the norm of the calibrated magnetometer measurements as done in \Figureref{fig:normMagField} does not indicate that the ML estimate is to be preferred over the estimate from the initialization phase. Hence, analysis of the norm of the calibrated magnetometer measurements does not seem to be a sufficient analysis to determine the quality of the calibration in the case when the calibration is performed to improve the heading estimates.  

\section{Conclusions}
We have developed a practical algorithm to calibrate a magnetometer using inertial sensors. It calibrates the magnetometer for the presence of magnetic disturbances, for magnetometer sensor errors and for misalignment between the inertial and magnetometer sensor axes. The problem is formulated as an \ac{ml} problem. The algorithm is shown to perform well on real data collected with two different commercially available inertial measurement units. 

In future work the approach can be extended to include GPS measurements. In that case it is not necessary to assume that the acceleration is zero. The algorithm can hence be applied to a wider range of problems, like for instance the flight test example discussed in~\cite{kokHSGL:2012}. The computational cost of the algorithm would, however, increase, since to facilitate the inclusion of the GPS measurements, the state vector in the \ac{ekf} needs to be extended. 

Another interesting direction for future work would be to investigate ways of reducing the computational cost of the algorithm. The computational cost of the initialization steps is very small but actually solving the \ac{ml} problem in \Stepref{algstep:ml} of \Algorithmref{alg:ml} is computationally expensive. The algorithm both needs quite a large number of iterations and each iteration is fairly expensive due to the computation of the numerical gradients. Interesting lines of future work would either explore different optimization methods or different ways to obtain gradient estimates. 

Finally, it would be interesting to extend the work to online estimation of calibration parameters. This would allow for a slowly time-varying magnetic field and online processing of the data.

\section{Acknowledgements}
This work is supported by CADICS, a Linnaeus Center, and by the project \emph{Probabilistic modeling of dynamical systems}
  (Contract number: 621-2013-5524), both funded by the Swedish Research Council (VR), and by MC Impulse, a European Commission, FP7 research project. The authors would like to thank Laurens Slot, \dr Henk Luinge and \dr Jeroen Hol from Xsens Technologies and \dr Gustaf Hendeby from Link\"oping University for their support in collecting the data sets and for interesting discussions. The authors would also like to thank the reviewers for their constructive comments.

\newpage
\bibliographystyle{unsrtnat}
\bibliography{IEEEfull,literature}

\end{document}